\documentclass{sig-alt-release2}

\begin{document}

\conferenceinfo{JCDL Workshop on Collaborative Information Retrieval,} {June 20, 2008, Pittsburgh, Pennsylvania, USA.} 
\CopyrightYear{2008}
\crdata{978-1-59593-998-2/08/06}

\title{Toward Collaborative Information Seeking (CIS)}
\numberofauthors{1}
\author{
\alignauthor
{Chirag Shah}\\
       \affaddr{School of Information \& Library Science (SILS)}\\
       \affaddr{University of North Carolina}\\
       \affaddr{Chapel Hill NC 27599 USA}\\
       \email{chirag@unc.edu}
}

\maketitle
\begin{abstract}
It is natural for humans to collaborate while dealing with complex problems. In this article I consider this process of collaboration in the context of information seeking. The study and discussion presented here are driven by two dissatisfactions: (1) the majority of IR systems today do not facilitate collaboration directly, and (2) the concept of collaboration itself is not well-understood. I begin by probing the notion of collaboration and propose a model that helps us understand the requirements for a successful collaboration. A model of a Collaborative Information Seeking (CIS) environment is then rendered based on an extended model of information seeking.
\end{abstract}




\section{Introduction}
{\em Two is company; three is a crowd.} This seems like a simple saying, but going beyond a snippet of wisdom, it says a lot. The meaning of this proverb is obvious - {\em two heads are better than one}, but we have to be careful about having {\em too many cooks in the kitchen}. One of the objectives of the present article is to explore the idea of putting two or more heads together in a pursuit of a common goal. I will refer to this process as collaboration, the definition of which will be unveiled in Section 2. The other objective of this article is to map the notion of collaboration to the information seeking process. A model to evoke this discussion is proposed in Section 3. I conclude the article in Section 4 by summarizing some of the concepts presented here and presenting suggestions for the future research in this direction.

\section{Defining Collaboration}
I will start the discussion toward Collaborative Information Seeking (CIS) by reviewing the notion of collaboration. In the literature, we find people using the term `collaboration' in different senses. Moreover, it is often confused with coordination and cooperation. Looking up the dictionary definitions of these three terms does not help much in differentiating them clearly. Fortunately for us, people like Denning and Yaholkovsky \cite{Denning2008Getting}, and Taylor-Powell {\it et al.} \cite{Taylor-Powell1998Evaluating} have done extensive work on linking up these concepts. Based on their insights, I present a model of collaboration in Figure 1. This model consists of five layers, with collaboration encapsulating the others.

\begin{center}
\begin{figure}[htbp]
\centering
\includegraphics[scale=0.45]{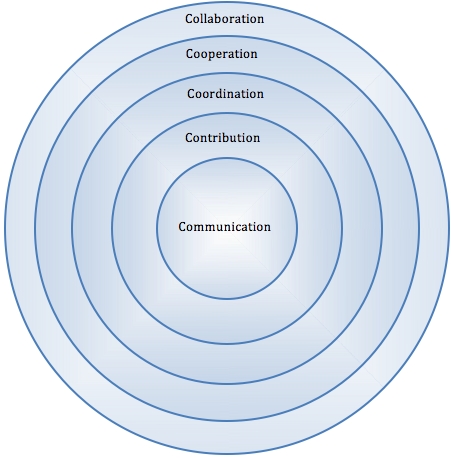} 
\caption{A model for collaboration}
\end{figure}
\end{center}
\vspace*{-1.4cm}

\begin{itemize}
\item{{\em Communication}. This is a process of sending or exchanging information, which is one of the core requirements for carrying out a collaboration, or maintaining any kind of productive relationship for that matter. For instance, Sarah putting up flyers for her college student body president campaign is a form of communication that she is doing with the college community.}
\item{{\em Contribution}. This is an informal relationship by which individuals help each other in achieving their individual goals. For instance, Mark chipping in for Sarah's campaign fund is a contribution.}
\item{{\em Coordination}. This is a process of connecting parties together for a harmonious action. This often involves bringing people under an umbrella at the same time and place. During this process, the involved parties may share resources, responsibilities, and goals. For instance, assembling a group of students for Sarah's campaign requires coordination.}
\item{{\em Cooperation}. This is a relationship in which parties with similar interests take part in planning activities, negotiating roles, and sharing resources to achieve joint goals. In addition to coordination, cooperation involves all the parties following some rules of interaction. For instance, during the campaign, another candidate, Michael, realizes that his and Sarah's campaigns are pretty much the same. So he decides to merge his campaign with that of Sarah. They negotiate their roles and responsibilities in this process with Sarah running for president and Michael for vice-president. They decide to cooperate, instead of compete.}
\item{{\em Collaboration}. This is a process of involving parties that may see different aspect of a problem. They engage in a process through which they can go beyond their own individual expertise and vision by constructively exploring their differences and searching for common solutions. In contrast to cooperation, collaboration involves creating a solution that is more than merely the sum of each party's contribution. The authority in such a process is vested in the collaborative rather than in an individual entity. For instance, after the election between Sarah's and Jose's parties (the only two parties remained in the race), none received the majority of the votes. They, in turn, decided to merge and create a collaborative. Both of these parties had slightly different visions and strengths, but now they have brought those together to create a shared solution that is likely to be different than what either of them could have achieved. In addition to this, we can hope that this solution is also a better one, since often a group of entities are found to create a much better solution than any individual entity by itself \cite{Surowiecki2004Wisdom}.  As Chrislip and Larson \cite{Chrislip1994Collaborative} define collaboration, ``It is a mutually beneficial relationship between two or more parties who work toward common goals by sharing responsibility, authority, and accountability for achieving results." Similarly, according to Gray \cite{Gray1989Collaborating:}, collaboration is ``a process through which parties who see different aspects of a problem can constructively explore their differences and search for solutions that go beyond their own limited vision of what is possible."}
\end{itemize}

Let us look back to the terms `coordination' and `cooperation', and see how they fit around this understanding of collaboration.   Austin and Baldwin \cite{Austin1991Faculty} note that while there are obvious similarities between cooperation and collaboration, the former involves pre-established interests, while the latter involves collectively defined goals. Malone \cite{Malone1988What} defined coordination as {\em the additional information processing performed when multiple, connected actors pursue goals that a single actor pursuing the same goals would not perform.}  While this definition is close to the one we have seen about collaboration, one can argue that it still fits in the model described in Figure 1 since it says nothing about creating solutions.

Another interesting observation we can make from the above mentioned model is that various terms are defined considering some form of relationship among people, and not between autonomous systems or humans and machines. In the literature, `collaboration' is also used when certain information such as queries and results are combined algorithmically \cite{Hust2005Query}. A stream of research, often referred to as {\em collaborative filtering} or {\em recommender systems} focuses on filtering the information presented to a user based his own history or the search behavior of the community around him \cite{Adomavicius2005Toward}. I will refrain from using this sense of collaboration and focus only on the kind of collaboration that is actively carried out among a group of people.

Let us now discuss how this understanding of collaboration informs the essentials of doing collaboration. In other words, let us explore the characteristics or the requirements of a successful collaboration. Surowiecki \cite{Surowiecki2004Wisdom} lists four such conditions:

\begin{enumerate}
\item{{\em Diversity of opinion}. Each person should have some private information, even if it is just an eccentric interpretation of known facts.}
\item{{\em Independence}. People's opinions are not determined by the opinions of those around them.}
\item{{\em Decentralization}. People are able to specialize and draw on local knowledge.}
\item{{\em Aggregation}. Some mechanism exists for turning private judgments into a collective decision.}
\end{enumerate}

All these conditions seem to make sense. We can also see what {\em cannot} be called a true collaboration. For instance, imagine two co-workers sitting in front of a computer seeking and analyzing some information. They may have diverse opinions, but they are not working independently in a decentralized way.\footnote{Note that decentralization does not necessarily imply that the co-workers are located remotely. Two co-workers can be co-located and can still work in a decentralized fashion.} What these conditions tell us is that in order to have a successful collaboration while seeking information, we need to create a supportive and environment where:

\begin{enumerate}
\item{The participants of a team come with different backgrounds and expertise.}
\item{The participants have opportunity of exploring information on their own without being influenced by the others, at least during a portion of the whole information seeking process.}
\item{The participants should be able to evaluate the discovered information without always consulting the others in the group.}
\item{There has to be a way to aggregate individual contributions to arrive at the collective goal.}
\end{enumerate}

One important aspect the above requirements is missing are the kind of task. There may not be much point in collaborating for simple fact-finding information tasks. As Morris and Horvitz \cite{Morris2007SearchTogether:} expressed, tasks that are exploratory in nature are likely to benefit from collaboration.

I will now take this understanding of the notion of collaboration and the requirements to have a successful collaboration to proposing a set of guidelines for a CIS environment. These guidelines, following the model in Figure 1, and derived from our discussion earlier, are given below.

\begin{enumerate}
\item{There has to be a way for the users of a CIS system to communicate with each other.}
\item{Each user should be able to (and encouraged to) make individual contributions to the collaborative.}
\item{User actions, information requests, and responses have to be coordinated to have an active and interactive collaboration. This collaboration could be synchronous or asynchronous, and co-located or remote.}
\item{Users need to agree to and follow a set of rules to carry out a productive collaboration. For instance, if they have a disagreement on the relevancy of an information object, they should discuss and negotiate; they should arrive to a mutually agreeable solution rather than continuing to dispute it. The system, of course, needs to support such a discussion and negotiation processes among the users.}
\item{The system should provide a mechanism to let the users not only explore their individual differences, but also negotiate roles and responsibilities. There may be a situation in which one user leads the group and others follow (cooperate), but the real strength of collaboration lies in having the authority vested in the collaborative, as noted earlier.}
\end{enumerate}

\section{A Model for Collaborative Information Seeking (CIS)}
Focusing our attention to information seeking now, we will see how our guidelines presented in the previous section can be realized in a model of information seeking. To begin our discussion, I present a four layer model of information seeking, centered around information access and organization in Figure 2. On the left side, four layers are labeled, on the right side, examples are given for these layers, and in the middle, a typical scenario is presented. These four layers are described below in detail.\\

\begin{center}
\begin{figure}[htbp]
\centering
\includegraphics[scale=0.5]{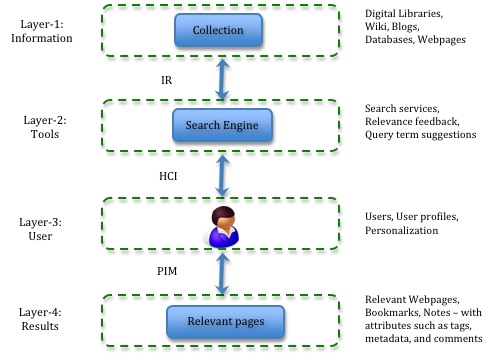} 
\caption{Four layer model of information seeking centered around information access and organization}
\end{figure}
\end{center}
\vspace*{-1cm}

\noindent
{\bf Layer-1: Information}\\
This layer contains information in various sources and formates (structured, semi-structured, and unstructured). The sources include digital libraries, wikis, blogs, databases, and webpages; formats include text, images, and videos.\\

\noindent
{\bf Layer-2: Tools}\\
This layer consists of tools and techniques a user can use to access the information of layer-1. They include search services, relevance feedback \cite{Buckley1994The-effect}, and query term suggestions \cite{anick03}. In addition, since this layer also acts as a mediating layer between information and users, it includes a variety of user interfaces, starting from results as rank-lists to touch panels with mechanisms to visualize results. We can see that a large amount of research in IR is focused on the link between layer-1 and layer-2; that is, developing tools and services appropriate for retrieving information of various forms.

\noindent
{\bf Layer-3: User}\\
This layer consists of a user, who uses the tools in layer-2 to access the information in layer-1 and accumulate the knowledge in layer-4. We can see that the focus of HCI research has been on the link between layer-2 and layer-3; that is, presenting the information and the information access tools in effective ways to the user. This layer-3 also includes elements relating to a user, such as user profiles, which can be used for personalization \cite{Teevan2005Personalizing}.\\

\noindent
{\bf Layer-4: Results}\\
The user of layer-3 accumulates the information relevant to him in layer-4. In the most basic sense, this could be a set of webpages that the user found relevant from his searches on the Web. Extending this further, we can have bookmarks, notes, and other kinds of results, sometimes stored with attributes such as tags, metadata, and comments. At a more conceptual level, this layer consists of the knowledge that the user gained by his information seeking process. The focus of research in PIM \cite{Dumais2003Stuff} has been on the link between layers 3 and 4, addressing the issues of information storage and organization by users.

\begin{center}
\begin{figure}[htbp]
\centering
\includegraphics[scale=0.36]{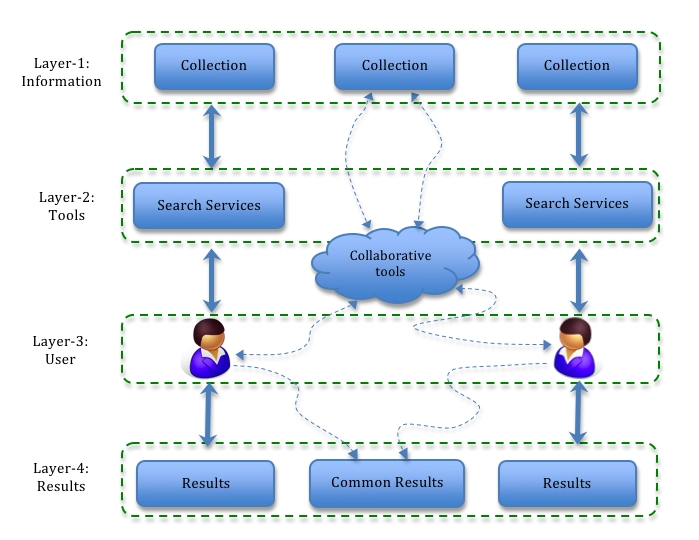} 
\caption{A model for Collaborative Information Seeking (CIS) extended from the model in Figure 2}
\end{figure}
\end{center}
\vspace*{-0.8cm}

I will now extend the general model of information seeking presented in the previous section to incorporate the notion of collaboration that I discussed earlier. To be specific, I will investigate the situations in which a group of users collaborate using traditional or collaborative tools to achieve personal or common information goals. A model with such a configuration is given in Figure 3. This is obtained by extending the original model of information access and organization for two users.\footnote{Considering only two users is merely for the simplicity; theoretically, the model can incorporate any number or users.} These users can access and organize information individually, or decide to collaborate with each other. In the case of collaboration, they will have a way to communicate with one another. They may have a common or shared interface. They may also have a shared space where they can store and organize their results. In other words, collaboration between these two users can occur at various levels: (1) while formulating an information request, (2) while obtaining results, and (3) while organizing and using the results.

\vspace*{-0.2cm}
\section{Conclusion}
In this article I attempted to formalize the notion of collaboration and proposed a model of Collaboration Information Seeking (CIS) that put collaboration in perspective. I presented a model of collaboration that incorporated often interchangeably used concepts, such as `coordination', `cooperation', and `collaboration'. With this model and related works on collaboration, I came up with requirements and guidelines for having a successful collaboration. This understanding was then translated into proposing a model for CIS.  I believe this model can not only help us implement CIS environments, but also evaluate existing systems for collaboration. Often ``collaborative" systems are designed by mixing various parts or objects (such as queries \cite{Fu2007Using} or documents \cite{Aslam2001Models}) of an information seeking process; however, I want to emphasize that a true collaborative system has to be based on a more user-centric design, giving the maximum freedom to the users, than letting the system carry out  all sorts of  ``collaboration" invisibly. My hope is to see CIS systems realized using a well-grounded model, such as the one proposed here, rather than simply extending a single-user IR system for multiple users.

Finally, one should also be cautious about the limitations of collaboration. It is important to point out that putting a group of people together does {\em not} always result in something that is {\em better} than what can be produced by an individual. Another significant aspect to consider is the kind of task. As noted earlier, collaboration may not prove to be effective in certain situations. All of these directions are worth investigating for a better understanding of CIS processes and enhancing user experience in collaborative environments.

\vspace*{-0.2cm}
\section{Acknowledgement}
The author is indebted to Gary Marchionini and Diane Kelly for their constant support and guidance on this work. He also wishes to thank Cassidy Sugimoto for reviewing this article and making important suggestions to improve it.
\newpage
\bibliographystyle{plain}

\end{document}